\documentstyle{article}
\begin{document}
\begin{titlepage}
\begin{flushright}
Z\"urich University Preprint\\
ZU-TH 18/95
\end{flushright}
\vfill
{\large\bf MICROLENSING RATES FROM
SELF-CONSISTENT GALACTIC MODELS}\\
\vfill
\centerline {F. De Paolis$^1$$^{\star}$, G. Ingrosso$^1$$^{\star}$,
Ph. Jetzer$^2$$^{\star\star}$}
\begin{center}{\it$^1$ Dipartimento di Fisica and INFN, Universit\`a di
Lecce,
CP 193, 73100 Lecce, Italy,\\
$^2$
Institute of Theoretical Physics, University of
Zurich,Winterthurerstrasse
190, CH-8057 Zurich, Switzerland}\\
\end{center}
\vfill
%\newpage
\begin{center}
Abstract
\end{center}
\begin{quote}
We study different models of dark matter distribution for the halo
of our galaxy. In particular, we consider Eddington and King-Michie
models, which include ani\-so\-tro\-py in the velocity space,
and compute in a self-consistent way the amount of
dark matter present in the halo.
Assuming that the dark matter
is in form of Massive Astrophysical Compact Halo Objects (MACHOs), we
find
for each model the expected number of microlensing events and their
average
time duration for an experiment monitoring stars in the
Large Magellanic Cloud (LMC).
The main effect of including anisotropy is to reduce the
microlensing rate by about 30\%
and to increase, but only slightly, the mean event duration,
as compared to the standard halo model.
Consideration of different luminous models for the visible part of the
galaxy also induce
variations in the microlensing results by roughly
the same amount as mentioned above.
The main uncertainty,
in order to be able to discriminate between different
dark matter distributions and to estimate the fraction of it
in form of MACHOs,
is due to the poor knowledge of the
rotation velocity at large
galactocentric distances.

\end{quote}
\vfill
$^{\star}$ Partially supported by Agenzia Spaziale Italiana.\\
$^{\star\star}$ Supported by the Swiss National Science Foundation.

\end{titlepage}
\newpage

\noindent{\bf 1. INTRODUCTION}\\

An important problem in astrophysics
lies in the knowledge of
the nature of the non-luminous matter present in galactic halos as
well as of their shape and extension.
The dark matter
in galactic halos may be in
the form of MACHOs (Massive Astrophysical Compact Halo Objects),
in the mass range $10^{-7} < M/M_{\odot} < \sim 10^{-1}$
(De R\'ujula et al. 1992)
, which can be
detected using the gravitational lensing effect (Paczy\'nski 1986).
%\cite{kn:Paczynski}.
The few microlensing events found so far by monitoring stars in the
Large Magellanic Cloud (LMC) by the EROS (Aubourg et al. 1993)
and the MACHO (Alcock et al. 1993)
collaborations,
while confirming the presence of MACHOs, do not yet allow to make
a precise estimate of the fraction of the halo dark matter in the
form of MACHOs nor to infer whether they are located in the halo or
rather
either in the LMC itself (Sahu 1994, Wu 1995, Gould et al. 1994)
or in a thick disk of our galaxy.
Assuming a standard spherical halo model, Alcock et al. (1995a)
found that MACHOs contribute a fraction $0.19^{+0.16}_{-0.10}$
to the halo dark matter, whereas their average mass turns out to
be $\sim 0.08 M_{\odot}$ (Jetzer 1994).
%\cite{kn:Jetzer}.

In this paper we investigate different models
of dark matter distribution for the halo of our galaxy.
In particular we consider the so-called Eddington and King-Michie
models,
which include anisotropy effects in the velocity space.
For these models we determine in a self-consistent way the amount
of dark matter present in the halo. Moreover,
assuming that the dark matter is in form of MACHOs we find for
each model the expected number of microlensing events
and their average time duration for an experiment monitoring stars
in the LMC.
These results are of relevance in order to determine
the fraction of the halo dark mass in form of MACHOs.
If the preliminary results of Alcock et al. (1995a)
mentioned above will
be confirmed by future improved observations, the problem
arises of how to explain the nature of the remaining fraction of the
halo dark matter. Besides the possibility of being composed
of new exotic particles, it could still be baryonic and in
form of molecular clouds (mainly $H_2$). This latter scenario
has been recently investigated in detail
by us in several papers (De Paolis et al. 1995, 1995a) and
also by Gerhard and Silk (1995).

In section 2 we describe the models for the dark matter distribution
and in section 3 we give the formulas for computing
the microlensing rate, the average time duration of an event
and the average mass of the MACHOs using the moment method.
The numerical results and the conclusions are given in section 4
and 5, respectively.\\

\noindent{\bf 2. MODELS FOR DARK MATTER DISTRIBUTION}\\

We assume that the galaxy contains two main components, namely, the
visible component and the dark component. In the following
we will assume that the dark matter consists
of MACHOs, although the results on the total amount of the dynamical
mass
do not depend on this assumption
and are, therefore, of more general validity.
In the last few years, the picture of the visible component of the
Milky Way
has evolved from the BSS model (Bahcall et al. l983)
which assumed stars to be distributed according to a central bulge,
a spheroid and an exponential disk.
In particular, the central concentration of stars is now described by a
triaxial bulge model with the density law
\begin{equation}
\rho_C(x,y,z) =\frac{M_b}{8\pi \tilde abc}e^{-s^2/2}~,~~~{\rm with}~~~
s^4=(x^2/\tilde a^2+y^2/b^2)^2+z^4/c^4~,
\label{eq:2.1}
\end{equation}
where the bulge mass is $M_b \sim 2 \times 10^{10}~M_{\odot}$ and
the scale lengths are $\tilde a=1.49$ kpc, $b=0.58$ kpc, $c=0.40$ kpc
(Dwek et al. 1994).
The coordinates $x$ and $y$
span the galactic disk plane, whereas  $z$ is perpendicular to it.
The remaining luminous matter
(the spheroid and the disk of the BSS model)
can be described with a double exponential disk
(see Gilmore et al. l989),
so that
the galactic disk has both a ``thin'' ($D_1$) and a ``thick'' ($D_2$)
component.
For the ``thin'' luminous disk we adopt the following
density distribution
\begin{equation}
\rho_D(X,z) = \frac {\Sigma_0 } {2H} ~e^{-|z|/H}~e^{-(X-R_0)/h},
\label{eq:2.2}
\end{equation}
where the local projected mass density is $\Sigma_0 \sim 25~M_{\odot}$
pc$^{-2}$, the scale parameters are $H\sim 0.30$ kpc and $h\sim 3.5$
kpc
and $R_0$=8.5 kpc is the local galactocentric distance.
Here $X$ is the galactocentric distance in the plane.
For the ``thick'' component we consider the same density law as in
eq.(\ref{eq:2.2}) but with variable thicknesses in the range
$H=1 \pm 0.5$ kpc
and local projected density
$\Sigma_0 \sim 50 \pm 25~M_{\odot}$ pc$^{-2}$.

The total local projected mass density within a distance of $(0.3-1.1)$
kpc of
the galactic plane is measured to be in the range
$(40-85)~M_{\odot}$ pc$^{-2}$.
This explains our chosen range of values for the ``thick'' disk, which
in fact
corresponds to a total luminous projected mass density of
$(50-100)~M_{\odot}$ pc$^{-2}$.
In our models we also consider the effect of varying
the bulge mass
$M_b = 2 \pm 1 \times 10^{10}~M_{\odot}$,
the local galactocentric
distance $R_0 = 8.5 \pm 1$ kpc and the scale length $h\sim 3.5 \pm 0.5$
kpc,
while we keep all the remaining parameters appearing
in the previous equations fixed.

We treat MACHOs composing the dark matter with the formalism based on
the
equation of state assuming that they are spherically symmetric
distributed
and that: \\
i) the velocities of MACHOs at any point of the galaxy are limited by
the escape velocity $v_c^2(r)$ from the galaxy itself;\\
ii) the distribution function can be anisotropic in the velocity space
as a consequence of the initial conditions of the galaxy formation
as well as of the MACHO formation processes
\footnote {It is well known that for elliptical galaxies models
in which the luminous matter is described by distribution functions
with
anisotropy in velocity space can account for their shapes. Moreover
both isotropic and anisotropic models can explain optical and X-ray
observations (De Paolis et al. 1995b).}.
Then, for MACHOs of the same mass $M$, we adopt the King-Michie
distribution
function  (King 1966, Michie 1963), which can be written as
(Ingrosso et al. 1992)
\begin{equation}
dn(r)= A (2 \pi \sigma^2)^{-3/2}~
e^{[W(r)-W(0)]}~ (e^{-v^2/2\sigma^2}-e^{-W(r)}) ~ e^{-L^2(r)/2 L^2_c}~
d^3 v~,
\label{eq:2.5}
\end{equation}
for $ v \le v_c(r)$ and $dn=0$ otherwise.
Here $W(r)=v^2_c(r)/2\sigma^2$ is the energy cutoff parameter,
$\bf L= \bf r \times \bf p$ the angular momentum and
$L_c$ the angular momentum cutoff.
Moreover $A$ is a normalization constant and $\sigma$ a parameter which
in the
limit of the classical statistics ($W(r) \rightarrow \infty$ and
$L_c \rightarrow \infty$) represents the one-dimensional MACHO velocity
dispersion.
It is clear that the above distribution function gives lower values for
the
phase space density at large values of energy and angular momentum as
compared
to the ones obtained from the Boltzmann distribution. Furthermore it
introduces an anisotropy in the velocity space, which increases
with the radial coordinate $r$ and leads to highly eccentric
orbits for the MACHOs located in the outer regions of the galaxy.

Visible and dark components are considered to be in hydrostatic
equilibrium
in the overall gravitational potential $V$ solution of the Poisson
equation
$\nabla^2 V = - 4 \pi G (\rho_H + \rho_C + \rho_{D_1} + \rho_{D_2})$,
which we solve assuming, as stated, spherical symmetry for the dark
mass
distribution.
Clearly, this assumption as well as the purely radial anisotropy in the
phase space for MACHOs are first order approximations, since the
presence
of the ``thin'' and ``thick'' stellar disks distort both the density
and
the velocity distribution of MACHOs close to the disk.
However, we are interested mainly in the microlensing rate expected
looking
towards the LMC and we have verified that the errors in the number of
events due to MACHOs close to the disks (i.e. in the region where
the errors due to our approximations are higher) are negligible respect
to those due to the uncertainties in the determination of the visible
component.

The distribution function given in eq.(\ref{eq:2.5}) is the most
natural
way to build self-consistent galactic models,
since the usual Boltzmann statistics
requires an external cutoff (in the density) to avoid the mass
divergence.
Vice versa eq.(\ref{eq:2.5}) naturally
implies a MACHO mass density equal to zero
at the boundary $R$ of the galaxy where $W(R)=0$.
Starting from eq.(\ref{eq:2.5}) and expressing $L_c = M r_a \sigma$
in terms of the anisotropy radius $r_a$ (Binney \& Tremaine 1987),
%\cite{kn:binney}),
we obtain, after an integration over $d^3 v$, the radial component for
the
mass density
%\cite{kn:ingrosso})
\begin{equation}
\rho_H(r)= A~ (2 \pi \sigma^2)^{-3/2}~ e^{[W(r)-W(0)]}~
\left(\frac{r_a}{r}\right)~
\int_0^{W(r)}[e^{-\xi}-e^{-W(r)}]~F(\lambda)d\xi ~,
\label{eq:2.6}
\end{equation}
where $\lambda=(r/r_a)\sqrt{\xi}$, $~\xi=v^2/2\sigma^2$,
$F(\lambda)$ is the Dawson integral (Abramowitz \& Stegun 1965)
%\cite{kn:abramowitz})
and $W(r)=-V(r)/\sigma^2$.

The distribution function in eq.(\ref{eq:2.5}) can be approximated
in the limit without energy cutoff ($W \rightarrow \infty$)
with a well treatable analytical model. The so-called Eddington model
(Binney \& Tremaine l987)
%\cite{kn:binney})
for which
\begin{equation}
dn(r)= A (2 \pi \sigma^2)^{-3/2}~
e^{[W(r)-W(0)]}~ e^{-v^2/2\sigma^2} ~ e^{-L^2(r)/2 L^2_c}~d^3 v~.
\label{eq:2.7}
\end{equation}
In this case one can easily show that for values of $r$ greater than
the
halo core radius $a$ (the region in which the galactic dynamical mass
is
dominated by the halo dark matter), the dark mass density $\rho_H(r)$
can be
approximated by
(Binney \& Tremaine 1987)
\begin{equation}
\rho_H(r)=\rho_0 \left(\frac{a^2+R_0^2}{a^2+r^2}\right)
\frac{1}{1+(r/r_a)^2}~,
\label{eq:2.8}
\end{equation}
where $\rho_0$ is the local dark mass density.

Before going on with the presentation of the numerical results for
the dark matter distributions, we discuss how we compute the
microlensing rate and the average time duration.\\

\noindent{\bf 3. MICROLENSING RATES AND MASS MOMENTS}\\

When a MACHO of mass $M$ is sufficiently close to the line of
sight between us and a
star in the LMC, the light from the source suffers a gravitational
deflection and the original star brightness increases by
\begin{equation}
A=\frac{u^2+2}{u(u^2+4)^{1/2}}~ .
\label{eq:bb}
\end{equation}
Here $u=d/R_E$ ($d$ is the distance of the MACHO from the line of
sight)
and $R_E$ is the Einstein radius defined as:
\begin{equation}
R_E^2=\frac{4GMD}{c^2}\tilde x(1-\tilde x)~,
\end{equation}
with $\tilde x=s/D$,
where $D$ and $s$ are the distance between the source (a star in the
LMC),
respectively the MACHO and
the observer.

Microlensing rates depend on the mass and velocity distribution of
MACHOs.
The mass density at a distance $s$
from the observer is given by
eq.(\ref{eq:2.6}) or (\ref{eq:2.8}).
However, the adopted halo model does not determine the MACHO number
density
as a function of mass. A simplifying  assumption is to let the mass
distribution be independent of the position in the galactic halo, i.e.,
we
assume the following factorized form for the number density per unit
mass
$dn/dM$,
\begin{equation}
\frac{dn}{dM}dM=\frac{dn_0}{d\mu} \frac{\rho_H(\tilde x)}{\rho_0}
d\mu=\frac{dn_0}{d\mu} H(\tilde x) d\mu~,
\label{eq:3.1}
\end{equation}
with $\mu=M/M_{\odot}$,
$n_0(\mu)$
not depending on $\tilde x$
and is subject to the normalization
$\int d\mu \frac{dn_0}{d\mu}M=\rho_0$.
Nothing a priori is known about the distribution $d n_0/d\mu$.

A different situation arises for the MACHOs velocity distribution,
whose
projection in the plane perpendicular to the line of sight
$f(\tilde x,v_T)$ is obtained
from eq.(\ref{eq:2.5}), by adopting cylindrical coordinates along the
microlensing tube (so that $d^3 v= v_T dv_T dv_x d\theta$).
After integration over the longitudinal velocity $v_x$ and the
direction
$\theta$ of the transverse velocity $v_T$ in the perpendicular plane
we get
\begin{equation}
\begin{array}{l}
f(\tilde x,v_T) =  (2 \pi \sigma^2)^{-3/2}~A~e^{[W(r)-W(0)]}~
\int_{-v_c(r)}^{+v_c(r)}~dv_x~[e^{-(v_x^2+v_T^2)/2\sigma^2}-e^{-W(r)}]
\nonumber \\  \\
\int_0^{2\pi}~d\theta~
e^{-(r/r_a)^2[(1-sin^2\gamma cos^2\theta)v_T^2/2\sigma^2
+sin^2\gamma v_x^2/2\sigma^2- sin2\gamma cos\theta v_x
v_T/2\sigma^2]}~,
\nonumber \end{array}
\label{eq:3.2}
\end{equation}
where $\gamma$ is the angle between the MACHO radial distance $\bf r$
and the distance $\bf s$ from the observer.
The average value $<v_T^m>$ is obtained by a further integration
\begin{equation}
<v_T^m(\tilde x)> = \int_0^{+v_c(r)} f(\tilde x,v_T) v_T^{(m+1)}
dv_T~.
\label{eq:3.3}
\end{equation}

In the case of anisotropic models with $W \rightarrow \infty$
(the Eddington model in eq.(\ref{eq:2.7}))
the previous equation can be cast into the following form:
\begin{equation}
<v_T^m(\tilde x)>=
\frac{\sqrt 2 \sigma^m} {[1+(r/r_a)^2 ~{\rm sin}^2\gamma]^{1/2}}~
\int_0^\infty I_0[\eta^2 g(\tilde x)]~e^{-\eta^2
[1+(r/r_a)^2+g(\tilde x)]}~\eta^{(m+1)}
d\eta~,
\label{eq:3.4}
\end{equation}
where $I_0$ is a Bessel function and $g(\tilde x)$ is given by
\begin{equation}
g(\tilde x) = \frac { (r/r_a)^2 [1+(r/r_a)^2] ~{\rm sin}^2\gamma}
	     { [1+(r/r_a)^2 ~{\rm sin}^2\gamma]}~.
\label{eq:3.6}
\end{equation}

In order to find the rate at which a single star
is microlensed with amplification
$A \geq A_{min}$ (or $u \leq u_{max}$), we consider MACHOs
with masses between $\mu$ and $\mu+ d\mu$, located at a distance from
the observer in the range $\tilde x$ and $\tilde x+ d\tilde x$
and with transverse velocity in the interval
$v_T$ and $v_T+dv_T$. The collision time can be
calculated using the well-known fact that the inverse of the collision
time is the product of the MACHO number density, the microlensing
cross-section and the average velocity.
The rate $d\Gamma$
at which a single star is microlensed
in the interval $d\mu d\tilde x dv_T$ is given
by (De R\'ujula et al. 1991, Griest 1991)
\begin{equation}
d\Gamma(\tilde x,\mu,v_T)  =
2 D r_E u_{max} v_T^2 f(\tilde x,v_T) [\mu \tilde x(1-\tilde x)]^{1/2}
H(\tilde x)
\frac{d n_0}{d\mu} d\mu d\tilde x dv_T,
\label{eq:zt}
\end{equation}
with
\begin{equation}
r_E= \left( \frac{4GM_{\odot}D}{c^2} \right)^{1/2} \sim
3.2\times 10^9~{\rm km}~.
\label{eq:zs}
\end{equation}

One has to integrate
the differential number of microlensing events,
$dN_{ev}=N_{\star} t_{obs} d\Gamma$,
over an appropriate range for $\mu$, $v_T$ and $\tilde x$,
in order to obtain the total number of microlensing events which can
be compared with an experiment
monitoring $N_{\star}$ stars during an
observation time $t_{obs}$ and which is able to detect
an amplification such that $A \geq A_{min}$.

The range of integration for
$v_T$ is $ 0 <v_T < v_c(\tilde x)$ for the King-Michie models, while
the upper limit goes to $\infty$ for the Eddington models. Moreover,
$\tilde x$ ranges between 0 and 1
and $\mu$ varies in the interval where the mass
distribution $dn_0/d\mu$ is not vanishing.

However, each experiment has time
thresholds $T_{min}$ and $T_{max}$ and only detects events with
$T_{min}\leq T \leq T_{max}$,
and thus the integration range has to be such that $T$ lies in this
interval.
The total number of micro-lensing events is then given by
\begin{equation}
N_{ev}=\int dN_{ev}~\Theta (T-T_{min})\Theta (T_{max}-T),
\label{eq:th}
\end{equation}
where the integration is over the full range of
$d\mu d\tilde x dv_T$ mentioned above. $T$ is related in a complicated
way
to the integration variables,
because of this no direct
analytical integration in eq.(\ref{eq:th}) can be performed.

In practice in order to evaluate eq.(\ref{eq:th}) we define
an efficiency function $\epsilon_0(\mu)$
which measures the fraction of the total number of microlensing events
that meet the condition on $T$ at a
fixed MACHO mass and thus takes into account
the $\Theta$ functions in eq.(\ref{eq:th}).
A more detailed analysis (De R\'ujula et al. 1991)
shows that $\epsilon_0(\mu)$
is, to a very good approximation, equal to unity for possible MACHO
objects
in the mass
range of interest.
We will therefore set $\epsilon_0(\mu)=1$ in the following.

{}From the
experimental lensing data it is possible to extract information on the
MACHO mass distribution (De R\'ujula et al. 1991).
%\cite{kn:Derujula2}.
It
proves most useful to average powers of the duration $T$, i.e., to
construct
``duration'' moments.
Consider an experiment that observes $N_{\star}$
stars during a time
$t_{obs}$,
that has recorded a set of microlensing events, each one with a
duration $T$. For each event we get
a value for the dimensionless variable $\tau$, defined as follows:
\begin{equation}
\tau = \frac{v_0}{r_E} T =\frac{v_0}{v_T}
[\mu \tilde x(1-\tilde x)]^{1/2}~,
\label{eq:ts}
\end{equation}
(here $v_0$ is the local rotational velocity)
and  we construct the
$n$-moment of $\tau$ from experimental data as:
$< \tau^n >= \sum_{events} \tau^n$.

The theoretical
expectations for these moments are calculated as follows:
\begin{equation}
< \tau^n >=\int dN_{ev}~\tau^n
= V \,u_{max}\,\, \gamma(m) <\!\mu^m\!>~ ,
\label{eq:wb}
\end{equation}
with $m \equiv (n+1)/2$ and
\begin{equation}
V \equiv 2N_\star t_{obs} D \, r_E \, v_0~,
%  = 26\,{\rm pc}^3\,\, \frac {N_\star}{10^6}
%  \, \frac {t_{obs}}{4\,{\rm months}},
\label{eq:we}
\end{equation}
\begin{equation}
\gamma(m) \equiv \int_0^{+v_c}
\int_0^1 d\tilde x \left(\frac{v_T} {v_0}\right)^{1-n} v_T
f(\tilde x,v_T)  [\tilde x(1-\tilde x)]^m H(\tilde x)~,
\label{eq:wf}
\end{equation}
\begin{equation}
<\! \mu^m \!> \, \equiv \int d\mu~\frac{dn_0}{d\mu} \mu^m.
\label{eq:wh}
\end{equation}

The main point of eq.(\ref{eq:wb}) is that the
experimental value of $<\tau^n>$ allows to determine the
moments of the MACHO mass distribution, which can be used to
reconstruct the
mass function itself.
The mean local density of MACHOs (number per cubic parsec) is
related to $< \mu^0 >$. The average local mass density of MACHOs is
proportional to $< \mu^1 >$ (solar masses per cubic parsec). The mean
MACHO
mass $\bar M$ is then given by
\begin{equation}
\bar M \equiv \frac{<\mu^1>}{<\mu^0>} = \frac{<\tau^1>}{<\tau^{-1}>}
\frac{\gamma(0)}{\gamma(1)}~.
\label{eq:wg}
\end{equation}
where $<\tau^1>$ and $<\tau^{-1}>$ are determined through the
observed microlensing events.
We may use
$< \mu^1 >$ to compute the fraction $f$ of the dark matter density
$\rho_0$
that has been detected in the form of MACHOs. Indeed, we have
\begin{equation}
f=\frac{M_{\odot}}{\rho_0}< \mu^1 >
\approx 126  < \mu^1 > ~ {\rm pc}^3~
\left(\frac {7.9\times 10^{-3}~M_{\odot}~{\rm pc}^{-3}}    {\rho_0}
\right)
{}~.\label{eq:wl}
\end{equation}
Finally, the event duration $T$ can be expressed in terms of the
$\gamma$
functions defined in eq.(\ref{eq:wf}) as follows:
\begin{equation}
T=\frac{r_E} {v_0} <\mu^1>^{1/2} \frac{\gamma(1)}{\gamma(1/2)}
{}~.
\label{eq:duration}
\end{equation}

In order to quantify the expected number of events it is convenient
to take as an example a delta function distribution for the mass, i.e.
$dn_0/d\mu=\delta(\mu-\bar\mu)/\mu$.
The rate of microlensing
events with
$A \geq A_{min}$ (or $u \leq u_{max}$), is then
\begin{equation}
\Gamma(A_{min})=\tilde\Gamma u_{max}=
2 D r_E u_{max} \frac{\rho_0}{M_{\odot}} \frac{1}{\sqrt{\bar \mu}}
\int_0^{+v_c} dv_T \int^1_0 d\tilde x~ v_T^2 f(\tilde x,v_T)
[\tilde x(1-\tilde x)]^{1/2} H(\tilde x)~.
\label{eq:ta}
\end{equation}

Assuming the standard halo model for the MACHO density distribution
(which one obtains by taking the limit
$r_a \rightarrow \infty$ in eq.(\ref{eq:2.8}))
$\rho_H(r)=\rho_0(a^2+R_0^2)/(r^2+a^2)$
with
$\rho_0 = 7.9 \times 10^{-3}~M_{\odot}~{\rm pc}^{-3}$, $a =5.5$ kpc,
using for the LMC distance
$D=50$ kpc and for the angle between the line of sight
and the direction of the centre of the galaxy $\alpha=82^0$,
we obtain (De R\'ujula et al. 1991, Jetzer 1991).
%\cite{kn:Derujula2},\cite{kn:Jetzer2}
\begin{equation}
\tilde\Gamma=4
\times 10^{-13}~\frac{1}{s}~\left( \frac{v_0}{215~{\rm km/s}}\right)
\left(\frac{1}{\sqrt{D/{\rm kpc}}}\right)
\left( \frac{\rho_0}{7.9 \times 10^{-3}~M_{\odot}~{\rm pc}^{-3}}\right)
\frac{1}{\sqrt{M/M_{\odot}}}\ .
\label{eq:tb}
\end{equation}
For an experiment monitoring $N_{\star}$ stars during an
observation time $t_{obs}$ the total number of events with a
amplification $A \geq A_{min}$ is:
$N_{ev}(A_{min})=N_{\star} t_{obs} \Gamma(A_{min})$.
In Table 1
we show some values of $N_{ev}$ for the LMC,
taking
$t_{obs}=1$ year, $N_{\star}=10^6$ stars and
$A_{min}$ = 1.34. \\

\noindent{\bf 4. RESULTS}\\

We first discuss the Eddington model whose results coincide, in the
isotropic
limit, with those given in Table 1 for the galactic
standard halo with flat rotation curve up to the LMC.
We get a set of Eddington models
by varying the values
of the parameters $a$, $\rho_0$ and $r_a$.
The amount of luminous matter, as discussed in section 2,
is in the range between
$4.8 \times 10^{10}~M_{\odot}$ (``minimum disk model'') and
$1.1 \times 10^{11}~M_{\odot}$ (``maximum disk model'').

For each set of parameters we properly compute the resulting rotation
curve
considering the circular speed of the exponential disks
(Binney \& Tremaine 1987) and the contribution due to the bulge and
dark halo.
A given model is physically acceptable
only if it leads to a flat rotation curve up to the distance
of the LMC (with a local
rotational velocity $v_0 =215 \pm 10$ km s$^{-1}$). In order to check
this requirement we perform a $\chi^2_{~LMC}$ test, which gives a
measure
of the flatness of the rotation curve in the region
$5~{\rm kpc}<r < 50~{\rm kpc}$.
This way we find a range for the parameters for which the
corresponding models satisfy the $\chi^2_{~LMC} < 1$ test.

In Table 2 we report the mean values of the allowed range for
the parameters $a$, $\rho_0$ and $r_a$.
The errors quoted in the Table are such that
$\sim 80\%$ of the allowed models, which fulfill the $\chi^2_{~LMC} <
1$ test,
have parameters whose values lie in the range defined
by the errors around the mean quantity.
Table 2 gives also the number of microlensing events $N_{ev}$ (as in
Table 1
we consider an experiment monitoring $10^6$ stars during 1 year and
for the MACHO we assume a mass of $10^{-1} M_{\odot}$),
the average event
duration $T$, the ratio $[\gamma(0)/\gamma(1)]$, which is related to
the mean MACHO mass $\bar M$, as well as the amounts
of dark mass $M_H^{~LMC}$ and $M_H^{~Tot}$ which are
inside the distance to the LMC and 250 kpc, respectively.
The latter value is of course arbitrary and should be regarded
as an illustration, given also the fact that the true extent
of the halo is unknown.

As can be seen from Table 2, the main effect of the anisotropy in
velocity
space
(second line) is to reduce the amount of $M_H^{~Tot}$
substantially. The dark
mass $M_H^{~LMC}$ also decreases with increasing anisotropy, however
by less than 20\%. This is due to the fact that
the anisotropy in the velocity space becomes particularly important
starting from a distance $r>r_a$ (see eq.(\ref{eq:2.8})).

The decrease of $M_H^{~LMC}$ for anisotropic models compared to
isotropic ones
implies a smaller number of microlensing events
$N_{ev}$, since the MACHOs orbits are now more eccentric.
This reduction is also partially due to a shortening of the
average transverse MACHO velocity $<v_T>$ with increasing anisotropy
which,
on the other hand, leads to
an increase of the event duration.
Finally, as one can see from the last column of Table 2, for the
anisotropic models there is a decrease
of the ratio $[\gamma(0)/\gamma(1)]$
leading (by eq.(\ref{eq:wg})) to a smaller average MACHO mass $\bar
M$.

Till now we assumed a flat rotation curve in the range $(5-50)$ kpc.
Actually, the galactic rotation curve is well measured only in the
range
$(5-20)$ kpc (Merrifield 1992).
Thus we
consider Eddington models for which we relax the condition
of a flat rotation curve in the range $(20-50)$ kpc.
To select acceptable physical models we follow
the strategy outlined, e.g.,
in Gates et al. (1995).
We require that:
i) the local rotational velocity is $v_0 = 215 \pm 10$ km/s; $~$
ii) the total variation in $v_{rot}(r)$ in the range $5~{\rm kpc}<r<20$
kpc
is less than 14\%; $~$
iii) the rotational velocity $v_{rot}(LMC)$ at the LMC
is in the range  $(150-307)$ km s$^{-1}$.

The results obtained for these Eddington models are given in Table 3,
where the first row corresponds to isotropic models.
A comparison between Tables 2 and 3 shows that relaxing the requirement
that the rotation curve has to be flat at distances larger than
$\sim 20$ kpc does not
change much the previous results. The main effect seems to be a
larger range of variation for our results, although
in Table 3 there is a trend towards a decrease of both
$M_H^{~LMC}$ and $N_{ev}$.
The values in Table 3 also show
the effect of the anisotropy which increases the
event duration and decreases the $[\gamma(0)/\gamma(1)]$ ratio.
Notice that in both Table 2 and 3 the ``minimum disk model'' leads
to the higher value for the dark mass (columns 4 and 5), whereas
the ``maximum disk model'' correlates with the smaller amount
in the mentioned range around the mean quantity.

Finally, we consider King-Michie models, whose results are
reported in Table 4.
Here, again we adopt the same strategy as in Table 2, considering
only models which lead to a flat rotation curve up to
the LMC distance (performing again a $\chi^2_{~LMC} < 1$ test).

As before the numerical results depend on the
assumed amount of luminous matter, so we consider, as an illustration,
the two extreme
models corresponding to the
``minimum'' (rows 1 and 4) and the
``maximum'' (rows 3 and 6)
disk together with an intermediate model (rows 2 and 5).

The values in Table 4 for the King-Michie models confirm the result
that
the anisotropy in phase space decreases $N_{ev}$
and $\bar M$, while it increases $T$. However, the variation is not so
marked
as for the Eddington models in Table 2.
Moreover, contrary to Table 2, there is no correlation between
the values of $M_H^{~LMC}$ and $N_{ev}$.
This can be understood by noting that increasing the amount of luminous
matter (from rows 1 to 3 for the isotropic models,
and from rows 4 to 6 for the
anisotropic ones) the dark matter central density decreases while
the core radius $a$ increases. Therefore, the dark matter density is
higher
in the central region of the galaxy as well as in the outermost
regions,
whereas it decreases in the space in between the
Solar system and the LMC. As a result of this
the amount of microlensing events $N_{ev}$
will be smaller.

The results in Table 4 show also an increase of the total dark matter
with
increasing luminous matter (dominated by the disk).
The increase of dark matter with a corresponding
increase of luminous matter
can be understood by noting that the halo density as well as the
velocity
dispersion of the dark matter (randomly in orbit
about the common center of mass)
must be sufficiently high to avoid collapse into the disk.
In the following paragraph we give a qualitative argument
to show that this result is physically reasonable.

Consider, first, a disk with a few dark particles distributed in a
sphere.
This may be thought of as an inhomogeneous sphere which must finally
collapse
to the disk over a sufficiently long period (due to gravitational
instability). What will happen is that as the particle in the
``sphere''
orbits around the centre of mass it will repeatedly cross the disk.
While it
is sufficiently close to the disk the particle
will be attracted towards it and hence
the orbit will be pulled into the plane.
Now consider the sphere with only a very slight excess density in the
equatorial disk. The inhomogeneity will still cause collapse into the
disk
but in a very much longer period. Clearly, as the density ratio of halo
to
excess matter tends to unity the sphere will be stabilized, whereas
if it tends to
zero it will be progressively de-stabilized. In other words, the time
for
collapse is a function of the density ratio. It tends to infinity as
$\rho_H/\rho_{disk} \rightarrow 1$ and it tends to zero as
$\rho_H/\rho_{disk} \rightarrow 0$.

Now we turn our attention to the results in Table 4.
The larger scatter of the values in the Table compared to the ones in
Table 2
can be understood as being due to the variation of the amount of
luminous
matter as well as
to the fact that the King-Michie models have an
additional parameter (the central energy cutoff parameter $W_0$)
with respect to the Eddington models (for which $W_0 \rightarrow
\infty$).
Due to the presence of this additional parameter it is not possible
to see a clear correlation between the microlensing results
and the degree of anisotropy in velocity space, since the latter effect
can be smaller than the one induced by the variation of $W_0$. The
effect of
anisotropy would
show up clearly once we compare models which all have
a given value of $W_0$.

The difficulty of getting precise
microlensing results for the King-Michie
models becomes even more evident if we adopt
the same strategy to select physical models, which lead to Table 3,
that is by considering the conditions
i) - iii).
In Figure 1a we plot $N_{ev}$ as a function of the flatness
degree $[v_{rot}(LMC)/v_0]$ of the galactic rotation curve.
%One can see that the above parameter gives a
%measure of $N_{ev}$.
In Figure 1b, the same plot is given for the event duration $T$.
In both figures the external and internal ellipses define the regions
in
which the models satisfy the condition
of having a flat rotation
curve only up to 20 kpc and to the LMC distance, respectively.
It is clear that the knowledge of the LMC rotation velocity
is crucial for determining the number of microlensing events,
besides the ambiguity due to the uncertainty
of the amount of the galactic luminous matter. \\

\noindent{\bf 5. CONCLUSIONS}\\

Several authors have studied the problem of determining the
number of the expected microlensing events or, equivalently,
the optical depth to microlensing by considering different
models for the mass distribution, both luminous and dark
in the galaxy (see, e.g. Alcock et al. 1995b, Kerins 1995, Kan-ya 1995,
Evans
\& Jijina 1994, Evans 1994).
Alcock et al. (1995b)
considered ``power-law models''
for the halo and found that the microlensing rate can vary by as much
as
a factor 10 with respect to the value one gets for the standard halo.
Kan-ya et al. (1995)
analysed axisymmetric ``power-law
model'' and studied the variation of the optical depth to microlensing
and find that it can vary within a factor 2.5 compared to the standard
spherical halo model. Similar conclusions are reached by all
other authors.

The main point in our present work is that
we analyse another class of models (Eddington and King-Michie) which
have
anisotropy in the velocity space and we determine
the corresponding parameters
in a self-consistent way as described in section 2.
We find that with the present knowledge of the
various parameters the variation in the expected number
of microlensing events is at least within 30\%
from the value one gets for the standard halo model
(flat rotation curve up to the LMC).
This factor can even increase if one allows for less
restrictive conditions.
The typical time duration on the contrary seems to vary
less, as well as the ratio $[\gamma(0)/\gamma(1)]$ which is related
to the average MACHO mass. The determination of the latter
with the moment method seems thus to be quite robust, in fact the
ratio $[\gamma(0)/\gamma(1)]$, as compared to the value for
the standard halo model, varies by at most $\sim \pm 30\%$.
The main source of uncertainty, in order to be able to discriminate
between different dark matter distributions and to estimate the
fraction
of it in form of MACHOs is due to the poor knowledge of the
rotation velocity at large galactocentric distances
up to the LMC. \\

\vspace*{3cm}
\noindent {\bf ACKNOWLEDGEMENTS} \\
\noindent
GI and FD would like to thank A. Qadir
for reading the manuscript and for many useful suggestions.

\newpage
\vspace*{2cm}
\begin{center}
\begin{tabular}{|c|c|c|c|c|}\hline
MACHO mass in units of
$M_{\odot}$ & Mean $R_E$ in km & Mean microlensing time $T$ &
$N_{ev}$ \\
\hline
$10^{-1}$ & $3 \times 10^8$& 23.5 days & 5.6  \\
$10^{-2}$ & $10^8$         & 7.4 days  & 17.8   \\
$10^{-4}$ & $10^7$         & 17.8 hours & 178  \\
$10^{-6}$ & $10^6$         & 1.8 hours & 1784 \\
\hline
\end{tabular}
\end{center}

\vspace*{2cm}

\begin{center}
\begin{tabular}{|c|c|c|c|c|c|c|c|}
\hline
$a$&$\rho_0/10^{-3}$&$r_a$&$M_H^{~LMC}$&$M_H^{~Tot}$&
$N_{ev}$ & $T$ & $[\gamma(0)/\gamma(1)]$ \\
(kpc) & $(M_\odot~{\rm pc}^{-3})$ & (kpc)  & $(10^{11}~M_\odot)$ &
$(10^{11}~M_\odot)$ & &
(days) & \\
\hline
$6.3 \pm 1.4$ & $7.6 \pm 1.5$ & $\infty$ & $4.9 \pm 1.0$ & $26 \pm 5$ &
$5.4\pm 0.9$ &$ 22.4 \pm 0.7$ &$7.3 \pm 0.2$ \\
\hline
$6.5 \pm 1.4$ & $8.3 \pm 1.3$ &$47\pm 12$& $3.7 \pm 0.8$ & $7 \pm 2$ &
$4.1 \pm 0.6$ & $27.0\pm 1.1$ &$5.2\pm 0.3$ \\
\hline
\end{tabular}
\end{center}

\vspace*{2cm}

\begin{center}
\begin{tabular}{|c|c|c|c|c|c|c|c|}
\hline
$a$&$\rho_0/10^{-3}$&$r_a$&$M_H^{~LMC}$&$M_H^{~Tot}$&
$N_{ev}$ & $T$ & $[\gamma(0)/\gamma(1)]$ \\
(kpc) & $(M_\odot~{\rm pc}^{-3})$ & (kpc)  & $(10^{11}~M_\odot)$ &
$(10^{11}~M_\odot)$ & &
(days) & \\
\hline
$6.2\pm 1.4$ & $7.0 \pm 3.8$ & $\infty$ & $4.5 \pm 1.5$ & $24 \pm 9$ &
$4.7\pm 1.5$ &$ 22.5\pm 0.9$ & $7.7 \pm 0.2$ \\
\hline
$6.4\pm 1.4$ & $7.6 \pm 4.0$ &$40\pm 15$& $2.9 \pm 1.5$ & $5 \pm 2$ &
$3.5\pm 1.3$ & $27.7\pm 1.5$ & $5.7 \pm 0.3$ \\
\hline
\end{tabular}
\end{center}

\vspace*{2cm}

\begin{center}
\begin{tabular}{|c|c|c|c|c|c|c|c|}
\hline
$a$&$\rho_0/10^{-3}$&$r_a$&$M_H^{~LMC}$&$M_H^{~Tot}$&
$N_{ev}$ & $T$ & $[\gamma(0)/\gamma(1)]$ \\
(kpc) & $(M_\odot~{\rm pc}^{-3})$ & (kpc)  & $(10^{11}~M_\odot)$
& $(10^{11}~M_\odot)$ & &
(days) &  \\
\hline
$4.9 \pm 0.9$ & $9.1 \pm 0.9$ & $\infty$ & $6.0 \pm 0.9$ & $24 \pm 9$
&$6.4\pm 0.9$ &$ 22.6\pm 1.0$ & $8.2 \pm 0.8$ \\
$6.6 \pm 1.9$ & $6.4 \pm 0.7$ & $\infty$ & $6.3 \pm 0.9$ & $32 \pm 9$
&$5.9\pm 0.9$ &$ 21.7\pm 1.0$ & $8.5 \pm 1.0$ \\
$13  \pm 4  $ & $4.0 \pm 0.5$ & $\infty$ & $6.9 \pm 1.0$ & $49 \pm 9$
&$5.8\pm 0.9$ &$ 19.2\pm 1.0$ & $10.2 \pm 1.1$ \\
\hline
$5.7\pm 1.5$ & $9.7 \pm 1.1$ & $40\pm 15$ & $4.9 \pm 0.7$ & $11\pm 3$
&$5.7\pm 0.8$ & $25.1 \pm 0.9$ & $6.9 \pm 0.6$ \\
$8.9\pm 3.5$ & $7.0 \pm 1.0$ & $42\pm 16$ & $5.1 \pm 0.8$ & $14\pm 3$
&$5.2\pm 0.8$ & $24.1 \pm 1.3$ & $7.3 \pm 1.1$ \\
$17 \pm 3  $ & $4.5 \pm 0.6$ & $45\pm 13$ & $5.8 \pm 0.7$ & $21\pm 3$
&$5.0\pm 0.9$ & $21.6 \pm 1.0$ & $8.6 \pm 0.8$ \\
\hline
\end{tabular}
\end{center}

%\newpage
\noindent{\bf REFERENCES}\\

\noindent
%\bibitem{kn:abramowitz}
Abramowitz, M., \& Stegun, I.A., 1965, Handbook
of Mathematical Functions (New York: Dover)\\
%\bibitem{kn:Alcock}
Alcock, C., et al. 1993, Nature, 365, 621\\
%\bibitem{kn:Alcock1}
Alcock, C., et al. 1995a, Phys. Rev. Lett.,
74, 2867; and submitted to ApJ\\
%\bibitem{kn:AlcockI}
Alcock, C., et al., 1995b, ApJ, 449, 28 \\
%\bibitem{kn:Aubourg}
Aubourg, A., et al. 1993, Nature, 365, 623\\
%\bibitem{kn:bss}
Bahcall, J.N., Schmidt, M., \& Soneira, R.M., 1983,
ApJ, 265, 730\\
%\bibitem{kn:binney}
Binney, J.J., \& Tremaine, S.T., 1987,
Galactic Dynamics
(Princeton University Press)\\
%\bibitem{kn:Depaolis}
De Paolis, F., Ingrosso, G., Jetzer, Ph., \& Roncadelli, M., 1995,
Phys. Rev. Lett., 74, 14;  and A\&A, 295, 567;
Comments on Astrophys. 18, 87\\
%\bibitem
De Paolis, F., Ingrosso, G., Jetzer, Ph., Qadir, A., \&
Roncadelli, M., 1995a, A\&A, 299, 647\\
%\bibitem{kn:dis}
De Paolis, F., Ingrosso, G., \& Strafella, F., 1995b,
ApJ, 483, 83\\
%\bibitem{kn:Derujula1}
De R\'ujula, A., Jetzer, Ph., \& Mass\'o, E., 1992,
A\&A, 254, 99\\
%\bibitem{kn:Derujula2}
De R\'ujula, A., Jetzer, Ph., \& Mass\'o, E., 1991,
MNRAS, 250, 348\\
%\bibitem{kn:dwek}
Dwek, E., et al. 1994, NASA/GSFC Report (unpublished)\\
%\bibitem{kn:Evans}
Evans, N.W., \& Jijina, J., 1994,
MNRAS, 267, L21\\
%\bibitem{kn:Evans1}
Evans, N.W., 1994, MNRAS, 267, 333\\
%\bibitem{kn:gates}
Gates, E.I., Gyuk, G., \& Turner, M.S., 1995,
Phys. Rev. Lett., 74, 3724; and to appear Phys. Rev. D.\\
Gerhard, O. \& Silk, J., 1995, astro-ph 9509149\\
%\bibitem{kn:gwk}
Gilmore, G., Wyse, R.F.G., \& Kuijken, K., 1989,
Ann. Rev. Astron. Astrophys., 27, 555 \\
%\bibitem
Gould, A., Miralda-Escud\'e, J., \& Bahcall, J.N., 1994,
ApJ, 423, L105\\
%\bibitem{kn:Griest1}
Griest, K., 1991, ApJ, 366, 412\\
%\bibitem{kn:ingrosso}
Ingrosso, G., Merafina, M., Ruffini, R., \& Strafella, F., 1992,
A\&A, 258, 223\\
%\bibitem{kn:Jetzer2}
Jetzer, Ph., 1991, Atti del Colloquio di Matematica
(CERFIM) 7, 259\\
%\bibitem{kn:Jetzer}
Jetzer, Ph., 1994, ApJ, 432, L43\\
%\bibitem{kn:Kanya}
Kan-ya, Y., Nischi, R., \& Nakamura, T., 1995
Kyoto University preprint 1347, astro-ph 9505130\\
%\bibitem{kn:Kerins}
Kerins, E., 1995, MNRAS
in press, astro-ph 9406040\\
%\bibitem{kn:king}
King, I.R., 1966, Astron. J., 71, 64\\
%\bibitem{kn:merrifield}
Merrifield, M.R., 1992, Astron. J., 103, 1552 \\
%\bibitem{kn:michie}
Michie, R.W., 1963, MNRAS, 125, 127\\
%\bibitem{kn:Paczynski}
Paczy\'nski, B., 1986, ApJ, 304, 1\\
%\bibitem{kn:Sahu}
Sahu, K.C., 1994, Nature, 370, 275 \\
%\bibitem
Wu, X., 1995, ApJ, 435, 66 \\

%\end{thebibliography}

%\newpage
\noindent{\bf FIGURE CAPTIONS}\\

\noindent Figure 1a:
The number of expected microlensing events $N_{ev}$
is given as a function of the ratio $[v_{rot}(LMC)/v_0]$
for the King-Michie models.
We consider an experiment monitoring $10^6$ stars in the LMC during an
observation time of 1 year and we assume, as an indication, a MACHO
mass
of $10^{-1}~M_{\odot}$.
Models with a flat rotation curve up to the LMC lie inside the inner
ellipse (see also Table 4),
whereas the ones with flat rotation curve only up to 20 kpc
are bounded by the outer ellipse. \\

\noindent Figure 1b:
The same as in Figure 1a for the average event duration T (in days).

\newpage
\noindent{\bf TABLE CAPTIONS}\\

\noindent Table 1: Galactic standard halo model. \\

\noindent Table 2: Mean values of parameters for Eddington models
with flat rotation curves up to the LMC (the different parameters
are described in the text). \\

\noindent Table 3: Mean values of parameters for Eddington models
with flat rotation curves up to 20 kpc. \\

\noindent Table 4: Mean values of parameters for King-Michie
models with flat rotation curve up to the LMC.  \\

\end{document}